\documentclass[a4paper,11pt]{article}
\pdfoutput=1 

\usepackage{jheppub} 
\usepackage{graphics,graphicx}

\usepackage{color}
\usepackage{wrapfig}
\usepackage{tensor}
\usepackage{bm}
\usepackage{bbm}
\usepackage[normalem]{ulem}
\usepackage{pdfpages} 
\usepackage{comment} 
\usepackage{pdfrender,xcolor}
\usepackage{amsthm}
\usepackage{framed}
\usepackage{float}
\usepackage{sectsty}
\usepackage{comment}

\usepackage{slashed}  
\usepackage[force]{feynmp-auto}
\usepackage{fancyhdr}
\definecolor{shadecolor}{gray}{0.95}

\graphicspath{ {./Figs/} } 

\usepackage {hyperref}
\hypersetup{
    colorlinks=true,
    linkcolor=blue,
    urlcolor=blue,
    linktoc=all
}

\newcommand{\bd}{\mathbf}

\newcommand{\bb}{\mathbb}

\newcommand{\Hom}{\text{Hom}}

\newcommand{\bbz}{\bb{Z}}

\def\be{\begin{equation}}
\def\ee{\end{equation}}
\def\bsh{\begin{shaded}}
\def\esh{\end{shaded}} 
\def\bpm{\begin{pmatrix}}
\def\epm{\end{pmatrix}}

\title{Boundary criticality via gauging finite subgroups: a case study on the clock model}

\author{Lei Su}
\affiliation{Department of Physics, University of Chicago, Chicago, Illinois 60637, USA}

\abstract{
Gauging a finite Abelian normal subgroup $\Gamma$ of a nonanomalous 0-form symmetry $G$ of a theory in $(d+1)$D spacetime can yield an unconventional critical point if the original theory has a continuous transition where $\Gamma$ is completely spontaneously broken and if $G$ is a nontrivial extension of $G/\Gamma$ by $\Gamma$. The gauged theory has symmetry $G/\Gamma \times \hat{\Gamma}^{(d-1)}$, where $\hat{\Gamma}^{(d-1)}$ is the $(d-1)$-form dual symmetry of $\Gamma$, and a 't Hooft anomaly between them. Thus it can be viewed as a boundary of a topological phase protected by  $G/\Gamma \times \hat{\Gamma}^{(d-1)}$. The ordinary critical point, upon gauging, is mapped to a deconfined quantum critical point between two ordinary symmetry-breaking phases ($d =1$) or an unconventional quantum critical point between an ordinary symmetry-breaking phase and a topologically ordered phase ($d\ge 2$) associated with $G/\Gamma$ and $\hat{\Gamma}^{(d-1)}$, respectively. Order parameters and disorder parameters, before and after gauging, can be directly related.  As a concrete example, we gauge the $\bbz_2$ subgroup of $\bbz_4$ symmetry of a 4-state clock model on a 1D lattice and a 2D square lattice.  Since the symmetry of the clock model contains $D_8$, the dihedral group of order 8, we also analyze the anomaly structure which is similar to that in the compactified  $SU(2)$ gauge theory with $\theta =\pi$ in $(3+1)$D and its mixed gauge theory. The general case is also discussed. } 
 
 \makeatletter
\def\@fpheader{\relax}
\makeatother

\begin{document} 
\maketitle
\flushbottom  

\pagenumbering{arabic}

\section{Introduction} 
It is known that the boundary of a symmetry-protected topological (SPT) phase can have nontrivial 't Hooft anomalies that  impose strong constraints on the possible phases of the boundary theory and the allowed phases  should saturate the anomalies \cite{chen2013,  wen2013, vishwanath2013physics,    kapustin2014anomalous, kapustin2014symmetry, kapustin2014anomalies,  senthil2015symmetry,   witten2016fermion}.  In particular, a theory with a mixed 't Hooft anomaly between two symmetries cannot have a trivially gapped phase where both symmetries are preserved \cite{kapustin2014anomalous, gaiotto2015generalized}. For example, there is a mixed anomaly between time reversal symmetry and the center symmetry of  the $SU(2)$ Yang-Mills theory with a theta term at $\theta =\pi$, and the phase diagram is constrained by this anomaly \cite{gaiotto2017theta}.   However, it is not clear in general when the phase transition between two allowed phases is continuous. In this work, we study one simple scenario where such a continuous transition is guaranteed: gauging a finite subgroup of the symmetry associated with an ordinary Landau-Ginzburg phase transition. 

\begin{figure}[t]
\includegraphics[width=0.5\textwidth]{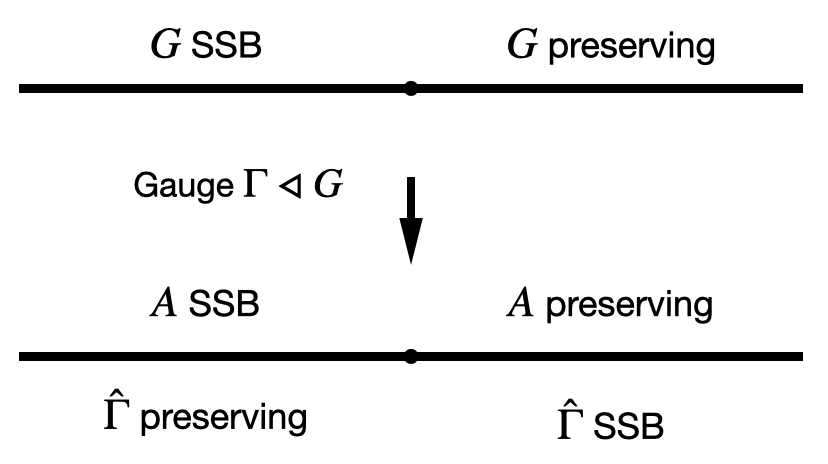}
\centering
\caption{Phase transition before and after gauging a finite Abelian normal subgroup $\Gamma$ 
of $G$. $A \equiv G/\Gamma$ and $\hat{\Gamma}^{(d-1)}$ is the dual symmetry of $\Gamma$ in the gauged theory. }
\label{phases}
\end{figure}

Suppose we have a theory $\cal{T}$ in $(d+1)$D spacetime with a nonanomalous 0-form symmetry $G$. We assume that $G$ has a nontrivial proper finite  Abelian normal  subgroup $\Gamma$. We can gauge $\Gamma$ of $G$ by coupling the theory to a flat background gauge field and then summing over flat background gauge fields with some weights \cite{Kapustin2014, gaiotto2015generalized}.
The gauged theory ${\cal{T}}/\Gamma$ has symmetry $G/\Gamma \times \hat{\Gamma}^{(d-1)}$, where $\hat{\Gamma}^{(d-1)}$ is the Pontryagin dual of $\Gamma$ \footnote{$\hat{\Gamma} = \Hom (\Gamma, U(1))$. For finite Abelian $\Gamma$ , $\hat{\Gamma}  \cong \Gamma$. Thus, in this work, we can use $\hat{\Gamma}$ and $\Gamma$ interchangeably.} and is a $(d-1)$-form symmetry whose charged objects are the original twisted sector operators \cite{ gaiotto2015generalized}. The mixed anomaly between $A \equiv G/\Gamma$ and $\hat{\Gamma}^{(d-1)}$ depends on the extension class $H^2(A, \Gamma)$ of the following sequence
\be 
1 \to \Gamma \to G \to A \to 1.
\label{seq0}
\ee
The observation was first applied to the $SU(2)$ gauge theory with $\theta =\pi$  in $(3+1)$D \cite{gaiotto2017theta} and 
the general proof is given in Ref.\cite{tachikawa2020} (see also Ref.\cite{wang2018}). If the mixed 't Hooft anomaly is nontrivial, disorder operators of $A$ are charged under $\hat{\Gamma}^{(d-1)}$ and vice versa.  Since gauging a finite subgroup is equivalent to coupling the theory to a discrete topological theory,  scaling laws of correlation functions of many operators are unaffected \cite{Kapustin2014}.  If the theory $\cal{T}$ has a continuous critical point where $\Gamma$ has a complete spontaneous symmetry breaking (SSB), then the gauged theory ${\cal{T}}/\Gamma$ has an unconventional quantum critical point associated with $A$ and $ \hat{\Gamma}^{(d-1)}$ if $G$ is a nontrivial extension of $A$ by $\Gamma$  (Fig.\ref{phases}). On one side, the $A$ is spontaneously broken but $\hat{\Gamma}$ is preserved. On the other side, the $\hat{\Gamma}$ is spontaneously broken but $A$ is preserved.   The critical behavior of the original theory shows up in the scaling laws of order operators and disorder operators of $G$. 
Upon gauging,  order operators and disorder operators of $A$ and $\hat{\Gamma}$ descend from those of $G$. The scaling laws across the critical point are also inherited. Thus, we have some exact information about the unconventional critical point from the ordinary critical point. 

\begin{table}[t]
\includegraphics[width=0.8\textwidth]{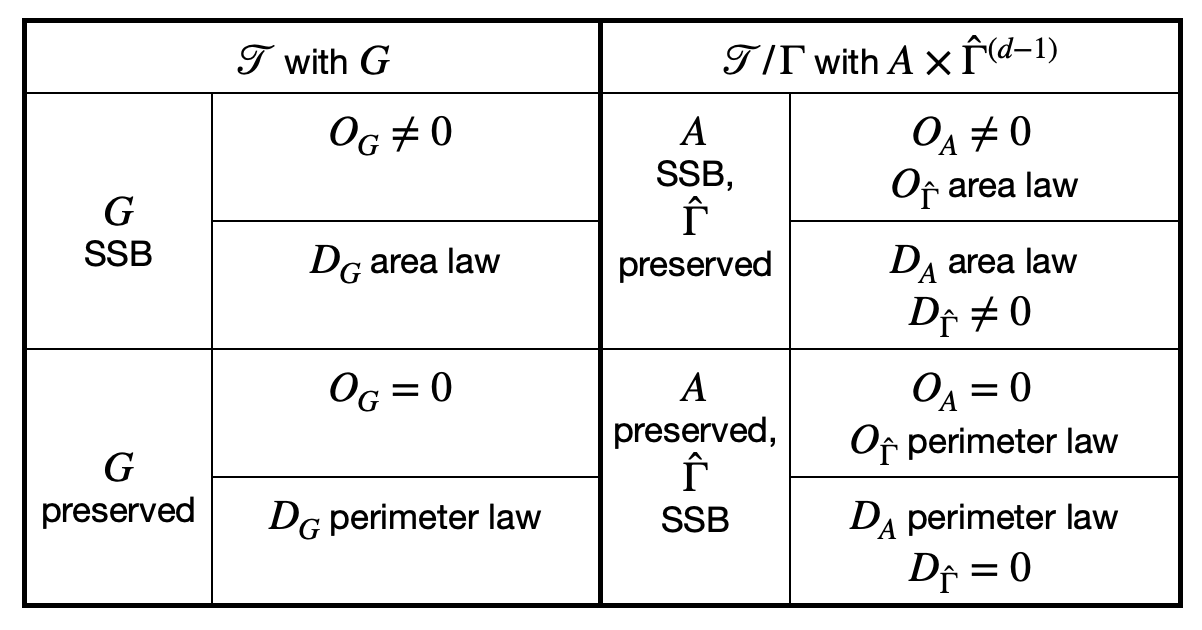}
\centering
\caption{Correspondences of scaling laws of order/disorder parameters in the ungauged theory $\cal{T}$ with symmetry $G$ and order/disorder parameters in the gauged theory ${\cal{T}}/\Gamma$ with symmetry $A \times \hat{\Gamma}^{(d-1)}$. We assume that the symmetry $G$ is completely broken across the critical point. When $d=1$, ``area law" should be taken to mean ``$=0$." }
\label{table1}
\end{table}

Mixed 't Hooft anomalies are naturally associated with deconfined quantum critical points (DQCP) if both symmetries $\hat{\Gamma}$ and $A$ are 0-form symmetries \cite{senthil2004deconfined, senthil2004quantum}. In fact, the essential idea we just described was used in Ref. \cite{zhang2023} to construct a DQCP in $(1+1)$D for $\Gamma =\bbz_2$ and $A =\bbz_2$ \footnote{The idea of gauging a $\bbz_2$ subgroup of $\bbz_4$ in the context of categorical symmetries was also mentioned in Ref.\cite{chatterjee2023symmetry, chatterjee2022holographic}.}. What we stated amounts to extending the idea to more general groups without stating the theory explicitly, i.e. obtaining DQCPs in $(1+1)$D by gauging a finite Abelian normal subgroup of a nonanomalous symmetry $G$ of a theory if the theory has a continuous SSB transition. Moreover, the idea naturally generalizes to higher dimensions where we necessarily have a 0-form symmetry $A$ and a higher-form symmetry $ \hat{\Gamma}^{(d-1)}$. Since spontaneously breaking a higher-form symmetry is associated with a topologically  ordered phase \cite{gaiotto2015generalized, mcgreevy2023generalized}, the phase transition is between an ordinary symmetry-breaking phase and a topologically ordered phase (enriched by $A$) .  

Order parameters and disorder parameters, before and after gauging, can be directly related.  For simplicity, let us consider the case when the symmetry $G$ of the original theory $\cal{T}$ is completely spontaneously broken, which can be detected by a single vanishing/non-vanishing order parameter $O_G$  order parameter \footnote{We use the two-point correlation function  associated with the order operator in the large distance limit as the order parameter. For example, in the Ising model, $O_G =\lim_{|i-j|\to \infty} \langle \sigma^z_i \sigma^z_j \rangle $.}. It can also be detected if the disorder parameter $D_G$  is vanishing/non-vanishing ($d=1$) or  satisfies the perimeter/area-law ($d\ge 2$).
Upon gauging, the order parameter $O_G$ of $\cal{T}$ becomes the disorder parameter $D_{\hat{\Gamma}}$ of symmetry $\hat{\Gamma}$ in gauged theory ${\cal{T}}/\Gamma$ but the disorder parameter $D_G$ of $\cal{T}$ remains the disorder parameter $D_A$ for the quotient symmetry $A$ in gauged theory ${\cal{T}}/\Gamma$. The order parameter $O_G$ of $\cal{T}$  can be viewed as  a (fractional) order parameter $O_A$ for the quotient symmetry $A$ in   ${\cal{T}}/\Gamma$ and   the disorder parameter $D_G$ of $\cal{T}$ can be viewed as a (fractional) order parameter  $O_{\hat{\Gamma}}$ of symmetry $\hat{\Gamma}$ \footnote{The end points of the string operators associated with $O_A$  are fractionally charged under $A$. See next section. }. Although gauging a finite subgroup changes the (non)locality of operators, it does not affect the scaling laws of these operators across the critical point. Thus, the scaling law of the order parameter $O_G$ in the ungauged theory $\cal{T}$ is the same as the scaling law of the disorder parameter $D_{\hat{\Gamma}}$ and   the scaling law of the  (fractional) order parameter $O_A$ in the gauged theory ${\cal{T}}/\Gamma$. Similarly, the scaling law of the disorder parameter $D_G$ in the ungauged theory $\cal{T}$ is the same as the scaling law of the disorder parameter $D_A$ and   the scaling law of the (fractional) order parameter   $O_{\hat{\Gamma}}$ in the gauged theory ${\cal{T}}/\Gamma$. 
These correspondences are summarized in Table.\ref{table1}.

The statements above may seem abstract but once we study a concrete example, we can see them immediately. With this mind, in this work we will work primarily with the 4-state clock model, whose total symmetry contains $\bbz_4$ or $D_8$, the dihedral group of order 8. The model has a continuous phase transition where $\bbz_4$ is completely broken. We will gauge the $\bbz_2$ subgroup of the symmetry group and study the mixed anomaly. We can eliminate the Gauss-law condition to obtain a simpler model where some symmetries act in a non-onsite way.  We discuss the $(1+1)$D case in section \ref{1D} and connect it to the 2D compact boson conformal field theory (CFT). We extend the analysis to $(2+1)$D in section \ref{2D}.  The connection to a continuum version of the bulk SPT in Ref.\cite{jian2021physics} was described in section  \ref{subsec_anomaly}. In addition, the similarity in the anomaly structure to the compactified $SU(2)$ gauge theory with $\theta =\pi$ in $(3+1)$D and its mixed gauge theory \cite{gaiotto2017theta} was analyzed in section \ref{subsec_mixed}. In section \ref{general}, the general case, especially $G =\bbz_n$, is discussed. We conclude with some discussion and future directions in the last section. Some detailed are relegated to appendices.

\section{Clock model in $(1+1)$D}
 \label{1D}
In this section, we consider $G= \bbz_4$ and $\Gamma  =\bbz_2$ in $(1+1)$D. We can start with any theory with $\bbz_4$ symmetry and gauge the $\bbz_2$ subgroup. Then as mentioned in the introduction, the gauged theory has  symmetry $\bbz_2 \times \bbz_2$, and the mixed anomaly is determined by the extension
 \be 
1 \to \bbz_2 \to \bbz_4 \to  \bbz_2 \to 1.
\label{seq1}
\ee
To be more concrete, we start with the 4-state clock model on a 1D chain as in Ref.\cite{zhang2023}: 
\be 
H  = -\sum_j (C^{\dagger}_j C_{j+1}  +  C^{\dagger}_{j+1} C_i) - g \sum_j (S_j +S_j^{\dagger}).
\label{clock1}
\ee
$C_j$ and $S_j$ are $4\times 4$ matrices that satisfy the relations
\be 
C_j^4  = S_j^4 = 1, \quad C_j S_j  = i S_j C_j.
\label{weyl}
\ee 
A periodic boundary condition is imposed. Apparently, it has a $\bbz_4$ symmetry generated by $U = \prod_j S_j$. The model is exactly solvable because it can be mapped to two decoupled Ising spin chains and the continuous critical point is at $g =1$. When $g>1$, the $\bbz_4$ symmetry is preserved; when $g<1$, the $\bbz_4$ symmetry is fully spontaneously broken. The order parameter can be taken to be  
$O =\lim_{|i-j|\gg 1} \langle C^{\dagger}_i C_j \rangle $ and the disorder parameter can be taken to be $D   =\lim_{|i-j|\gg 1} \langle \prod_{i\le k\le j}S_k\rangle $. In the disordered phase, the $O $ vanishes and $D  $ is a constant while in the ordered phase, $O $ is a constant but $D $ vanishes. At the critical point $g =1$, both $O $ and $D $ have a power law behavior $|i-j|^{-\eta}$ with $\eta =1/4$.

\subsection{Gauging $\bbz_2$}
\label{subsec_gauge1}
Now we gauge the $\bbz_2$ subgroup by coupling the model to Ising degrees of freedom $\tau^z_{i+1/2}$ on each link: 
 \be 
H  = - \sum_j (C^{\dagger}_j \tau^z_{j+1/2} C_{j+1}+  C^{\dagger}_{j+1} \tau^z_{j+1/2} C_j)  - g \sum_j (S_j +S_j^{\dagger}),
\label{gauged}
\ee
 with the Gauss-law condition on each site
 \be 
 \tau^x_{j-1/2} \tau^x_{j+1/2} = S_j^2.
 \label{constraints}
 \ee
 The gauged theory has two $\bbz_2$ symmetries:
 \be 
U_1  = \prod_j S_j, \quad U_2 = \prod_j \tau^z_{j+1/2}.
 \ee
Note that $U_1^2 =1$ because of the gauge constraints in Eq.( \ref{constraints}) and  that $U_2$ is the analogue of the dual $\bbz_2$ symmetry in the transverse-field Ising model after gauging the $\bbz_2$ symmetry or, equivalently, performing a Kramers-Wannier dual transformation. 

The disorder parameter $D =\lim_{|i-j|\to \infty} \langle \prod_{i\le k\le j}S_k\rangle $ is still gauge invariant. However, the order parameter $O =\lim_{|i-j|\gg 1} \langle C^{\dagger}_i C_j \rangle $ has to be dressed as  \be O =\lim_{|i-j| \gg 1} \langle C^{\dagger}_i \tau_{i+1/2}^z \cdots \tau_{j-1/2}^z C_j \rangle. \ee
Note that $  \prod_{i\le k\le j}S_k^2 =  \tau^x_{i-1/2} \tau^x_{j+1/2} $ implies that the domain walls associated with $U_1$ are fractionally charged under $U_2$. Similarly, the existence of the operators at the ends of the string operator $C^{\dagger}_i \tau_{i+1/2}^z \cdots \tau_{j-1/2}^z C_j$ implies that the domain walls associated with $U_2$ are fractionally charged under $U_1$. These are simply the consequences of the mixed anomaly between two $\bbz_2$ symmetries we mentioned above and lead to DQCPs \cite{levin2004deconfined}.

However, there is a loophole in this argument when generalizing it if the group extension is trivial. The new symmetries after gauging are still untangled. We give an example about gauging the $\bbz_2$ subgroup of the 6-state clock model in Appendix \ref{6clock}. Since mixed 't Hooft anomalies imply that not all symmetries can be realized in an onsite way \cite{wen2013}, to have a better sense of the mixed anomaly, we can eliminate the Gauss-law condition in Eq.(\ref{constraints}) to see if the symmetries are realized onsite.

\subsection{Eliminating the Gauss-law condition}
\label{map1}
To see the anomaly, we use the basis where $\tau^x_j$ and $S_j$ are diagonalized:
\be 
S  = \bpm 1 & 0 & 0& 0 \\ 0 & i & 0 & 0 \\ 0 & 0& -1 & 0\\ 0 & 0& 0& -i\epm, \quad C  = \bpm 0 & 1 & 0& 0 \\ 0 & 0 & 1 & 0 \\ 0 & 0&0 & 1\\ 1 & 0& 0& 0\epm.
\label{sc}
\ee
This is different from that in Ref.\cite{zhang2023} by a change of basis. 
\begin{figure}[tbh]
\includegraphics[width=0.8\textwidth]{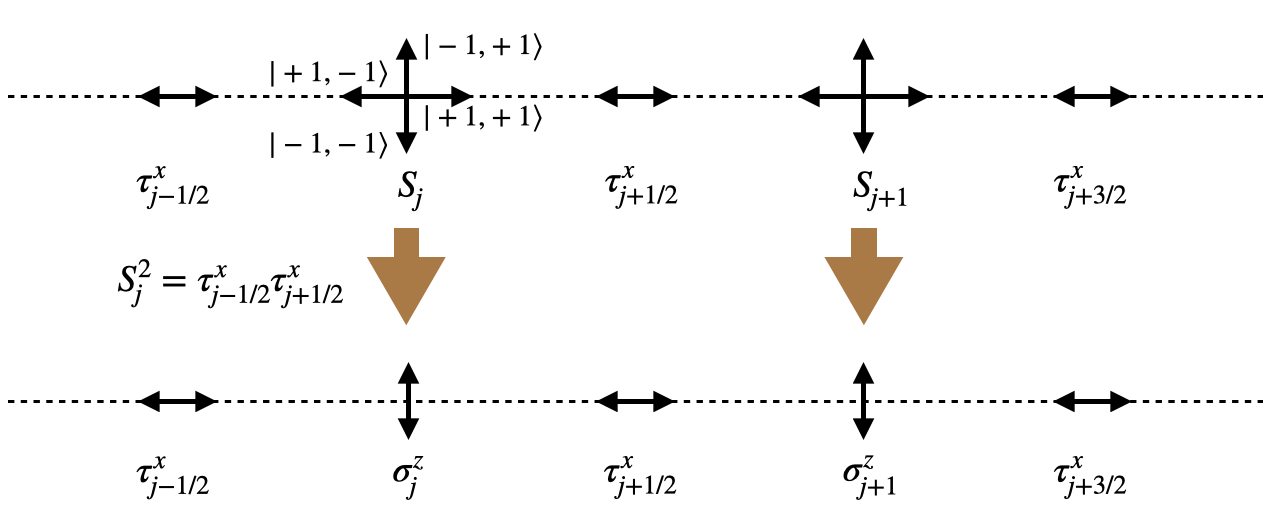}
\centering
\caption{Mapping of Hilbert spaces by eliminating the Gauss-law condition.  The second index in $|\pm 1, \pm 1\rangle$ is the eigenvalue of $\sigma_j^z$.}
\label{mapping}
\end{figure}

Note that due to the constraints in Eq.(\ref{constraints}) the states for  $\tau^x_j$ and for $S_j$ are not independent. We can reduce the local 4-dimensional Hilbert space on each site $j$ to 2-dimensional without losing information since $S_j^2$ is determined by the two neighboring $\tau^x_j$  (see Fig.\ref{mapping}). Namely, on each site $j$, we can rewrite  the basis state $|S_j \rangle$: $|S_j \rangle \to |\tau^x_{j-1/2}\tau^x_{j+1/2}, \sigma_j^z \rangle$ by the Gauss-law condition. The first index is the eigenvalue of $S_j^2$ while the second denotes the rest of degrees of freedom. It is easy to see that the constraints in Eq.(\ref{constraints}) implies

\be S_j \to \left(\frac{ 1+\tau_{j-1/2}^x \tau_{j+1/2}^x }{2}\right) \sigma_j^z + \left(\frac{ 1-\tau_{j-1/2}^x \tau_{j+1/2}^x }{2}\right) i \sigma_j^z.
\ee 
We can check that $S_j$ acting on the Hilbert space in the upper figure gives us the same eigenvalues on the Hilbert space in the lower figure. Naively, we may write 
\begin{equation*}
C_j^{\dagger}  \sim \left(\frac{1+ \tau_{j-1/2}^x \tau_{j+1/2}^x }{2}\right) \sigma_j^x + \left(\frac{1- \tau_{j-1/2}^x \tau_{j+1/2}^x}{2}\right)  ,
\end{equation*}
or 
\begin{equation*}
C_{j+1}   \sim\left(\frac{ 1+\tau_{j+1/2}^x \tau_{j+3/2}^x}{2}\right)\sigma_{j+1}^x  + \left(\frac{1- \tau_{j+1/2}^x \tau_{j+3/2}^x}{2}\right)    .
\end{equation*}
However, these two expressions are not exactly correct because action of $C_j^{\dagger} $ or $C_{j+1}$ alone violates the gauge-invariance. To fixed this problem, we must have

\be C_j^{\dagger} \tau_{j+1/2}^z  C_{j+1}  \to      K_j^{\dagger} \tau_{j+1/2}^z  K_{j+1},
\ee
where we have defined
\be 
K_j = \left(\frac{1+ \tau_{j-1/2}^x \tau_{j+1/2}^x }{2}\right) \sigma_j^x + \left(\frac{1- \tau_{j-1/2}^x \tau_{j+1/2}^x}{2}\right).  
\label{kk}
\ee 
Note that other than different notations  the mapping is essentially dual to that in Ref.\cite{zhang2023} ($ C_j^{\dagger} \tau_{j+1/2}^z  C_{j+1}\leftrightarrow S_j$) up to a basis rotation. Thus, 
\be 
U_1 =\prod_j S_j \to  U_a = \prod_j \sigma_j^z \ i^{\frac{ 1- \tau_{j-1/2}^x \tau_{j+1/2}^x}{2}}      , \quad U_2 =\prod_j C_j^{\dagger}  \tau_{j+1/2}^z   C_{j+1} \to   U_b = \prod_j \tau_{j+1/2}^z.
\ee
It can be checked that $U_a$ still generates a $\bbz_2$ symmetry because flipping a spin $\tau_{j-1/2}^x$  necessarily changes $\sum_j \tau_{j-1/2}^x \tau_{j+1/2}^x$ by a multiple of 4. Unsurprisingly, the action of $U_1$ is not onsite. We can further write down the reduced Hamiltonian explicitly:
\be 
H  =  - \sum (K_j^{\dagger} \tau_{j+1/2}^z  K_{j+1} +K_{j+1}^{\dagger} \tau_{j+1/2}^z  K_j) - g \sum (1+\tau_{j-1/2}^x \tau_{j+1/2}^x ) \sigma_j^z .
\label{redH}
\ee
This Hamiltonian can then describe an exactly solvable DQCP associated with $\hat{\Gamma}= \bbz_2$ and $A =\bbz_2$ as in Fig.\ref{phases}. The scaling laws of the order/disorder parameters are inherited from the original model. 
More details can be found in Ref.\cite{zhang2023}.
 
\subsection{$D_8$ symmetry}
\label{d8}
The 4-state clock model described by Eq.(\ref{clock1}) has more symmetries. In particular, it has the ``charge conjugation" symmetry ${\cal{C}} = \prod_j G_j$  with $G$ acting on the eigenstates $|s\rangle$ of $S$ as $G|s\rangle = |-s\rangle$, where $s =0,1,2,3 \mod 4$. In the matrix representation, it is given by
\be 
G  = \bpm 1 & 0 & 0& 0 \\ 0 & 0 & 0 & 1 \\ 0 & 0& 1 & 0\\ 0 & 1& 0& 0\epm. 
\ee
We can check that $G$ satisfies the following relations with $S$ and $C$ in Eq.(\ref{sc}):
\be
G S G =S^{\dagger}, \quad G C G =C^{\dagger}, \quad G^2 =1.
\ee
Thus the operator ${\cal{C}} = \prod_j G_j$ also generates a symmetry. Together with $U = \prod_j S_j$, it forms a larger non-Abelian group, the dihedral group $D_8$:
\be 
D_8 =\{{\cal{C}}, S| {\cal{C}}^2 =S^4=({\cal{C}} S)^2=1\}.
\ee
This non-Abelian polyhedral group is ubiquitous in the class of clock-like models (see Ref. \cite{cobanera2011bond} where what we called the clock model is named ``the quantum vector Potts model"). 

Even though $D_8$ is not completely broken in the clock model, gauging the center $\bbz_2$ symmetry produces a mixed anomaly nevertheless. Let us consider the effect of the gauging as in Eq.(\ref{gauged}) on ${\cal{C}}$. It is easy to see that ${\cal{C}}$ now commutes with both $U_1$ and $U_2$. The reason why ${\cal{C}}$ commutes with $U_1$  is that the  constraints  in Eq.(\ref{constraints}) force $ \prod_j S_j^2 =1$. This is not surprising because it simply the generalization of the extension in Eq.(\ref{seq1}) to 
 \be 
1 \to \bbz_2 \to D_8 \to  \bbz_2 \times \bbz_2 \to 1.
\label{seq2}
\ee
We can go through the same analysis in Section.\ref{map1} for ${\cal{C}}$. In terms of $\sigma_j^x$ and $\tau_{j-1/2}^x$, $G_j$ can be mapped to 
\be 
G_j \to G_j'= K_j \sigma_j^x,
\ee 
where $K_j$ is defined in Eq.(\ref{kk}),
and ${\cal{C}}$ is mapped to
\be 
{\cal{C}}' = \prod_j G_j' =\prod_j K_j \sigma_j^x.
\ee 
It is easy to see that $G_j'^2=1$, $G_j' S_j = S_j^{\dagger} G_j'   $. Thus, ${\cal{C}}'^2=1$ and  ${\cal{C}}' U_a  =U_a^3  {\cal{C}}' =U_a  {\cal{C}}'$. Also, since
${\cal{C}}' K_j^{\dagger} \tau_{j+1/2}^z  K_{j+1} {\cal{C}}'  =K_{j+1}^{\dagger} \tau_{j+1/2}^z  K_j  $,  ${\cal{C}}'$ is a symmetry of the reduced Hamiltonian in Eq.(\ref{redH}). In other words, $U_a$, $U_b$ and ${\cal{C}}'$ generate $\bbz_2^a \times \bbz_2^b \times \bbz_2^{\cal{C}'}$ of the reduced Hamiltonian. Both $U_a$ and   ${\cal{C}}'$  act in a non-onsite way. This is simply a consequence of the 't Hooft anomaly between $\bbz_2^b$ and $\bbz_2^a \times   \bbz_2^{\cal{C}'}$. On two sides of the critical point, either $\bbz_2^b$ is spontaneously broken or $\bbz_2^a \times   \bbz_2^{\cal{C}'}$ is spontaneously broken. In terms of the gauged theory, $\bbz_2^a$ and $\bbz_2^{\cal{C}'}$ are on an equal footing. More specifically, we are free to choose two of the three nontrivial order 2 elements as the generators and  call one of them the ``charge conjugation operator." This freedom plays a role in section \ref{subsec_mixed} when we compare the anomaly with the compactified $SU(2)$ gauge theory and its mixed theory. 

Note that  the $D_8$ symmetry of the 4-state clock model breaks to a $\bbz_2$ subgroup, as implied by the fact that there are only 4 degenerate ground states. Thus, $\bbz_2^a \times   \bbz_2^{\cal{C}'}$ also breaks to a $\bbz_2$ subgroup. In principle, we can construct models where $D_8$ is completely broken via a continuous critical point. For example, we may try to deform the 4-state clock model while preserving $D_8$.  In Ref.\cite{chatterjee2022holographic}, a model was constructed by coupling the 3-state Potts model to the Ising model. The theory has $S_3 = D_6$ and there is a continuous transition point where it is completely broken. Analogously, we may try to couple the 4-state clock model to the Ising model such that the total symmetry group is $D_8$ and there is a continuous transition point where it is completely broken.

\subsection{Connection to the compact boson CFT}
\label{sec_boson}
In the above discussion, we started with the 4-state clock model in Eq.(\ref{clock1}) which can be mapped to two decoupled Ising models. This is the ``hard" version of the clock model where degree of freedom at each site is strictly discretized and no $U(1)$ symmetry is emergent at the critical point, unlike the ``hard" $q$-state clock models when $q \ge 5$ \cite{patil2021unconventional}. However, if we use the ``soft" version,  the low energy physics of 4-state clock model after dropping marginally irrelevant cosine terms \cite{wiegmann1978one, li2020critical} 
 is the compact boson CFT \cite{francesco2012conformal}:
\be 
S = \frac{1}{4\pi} \int dz d\bar{z} \partial_z \phi \partial_{\bar{z}} \phi,
\ee
where $\phi(z, \bar{z}) \sim \phi(z, \bar{z}) +4\pi R$. The critical point of the 4-state clock model occurs at  $R=1$, as can be determined by the self-duality via the Kramers-Wannier transformation.  Let $\phi(z, \bar{z}) = X_L(z)+ X_R(\bar{z})$. Then the local primary operators with respect to the chiral algebra are the  vertex operators:
\be 
V_{n, w}(z, \bar{z} )= \exp\left[ i \left(\frac{n}{R} + w R\right) X_L(z)+i \left(\frac{ n}{R} -   w R\right) X_R(\bar{z}) \right],
\ee
where $n \in \frac{1}{2} \bbz$ and $w\in 2 \bbz$ are the momentum number and the winding number, respectively. \footnote{Here, we choose a convention for later convenience. We can define $n \in \bbz$ and $w\in \bbz$ by rescaling $R$ by a factor of 2.}   The conformal weights of $V_{n, w}$ are 
\be 
h_{n, w} = \frac{1}{4}  \left(\frac{n}{R} + w R\right) ^2, \quad \bar{h}_{n, w} = \frac{1}{4}  \left(\frac{n}{R} - w R\right) ^2,
\ee
and $\Delta_{n, w} =h_{n, w} +\bar{h}_{n, w}= (n^2/R^2 + w^2 R^2) /2$.
At a generic radius, the CFT has global symmetry $(U(1)_n \times U(1)_w) \rtimes \bbz_2^{\cal{C}}$ which acts as \cite{ji2020top}: 
\be 
U(1)_n: V_{n, w}  \to  e^{i 2 n\theta_n} V_{n, w}, \quad  
U(1)_w:  V_{n, w}  \to   e^{i w \theta_w/2} V_{n, w},\quad  
\bbz_2^{\cal{C}}: V_{n, w}  \to     V_{-n,- w} , 
\ee
where  $\theta_{n,w} \sim \theta_{n,w}+2\pi $. On $X_{L, R}$, the action is
\begin{align}
U(1)_n: & X_L(z)  \to X_L(z) +  R  \theta_n, \quad X_R(\bar{z}) \to X_R(\bar{z}) +  R  \theta_n \nonumber \\
U(1)_w: & X_L(z)  \to X_L(z) + \frac{1}{4R} \theta_w, \quad  X_R(\bar{z}) \to X_R(\bar{z}) - \frac{1}{4R} \theta_w \nonumber \\
\bbz_2^{\cal{C}}: & X_L(z) \to -X_L(z), \quad  X_R(\bar{z}) \to -X_R(\bar{z}).
\end{align} 
In our case, we focus on the $\bbz_4^n$ subgroup of $U(1)_n$ and consider the symmetry $D_8 = \bbz_4^n \rtimes \bbz_2^{\cal{C}}$ which act on $ V_{n, w} $ as
\be  
\bbz_4^n:    V_{n, w} \to i^{2n}  V_{n, w},  \quad
\bbz_2^{\cal{C}}:   V_{n, w} \to   V_{-n, -w}   .
\ee
There is also a T-duality $V_{n, w}(R) \to V_{w/4, 4n} (1/R)$ and a generalization of the Kramers-Wannier duality  \cite{ji2020top}.
We can gauge the $\bbz_2^n$ subgroup of $\bbz_4^n$
 \be 
1 \to \bbz_2^n \to \bbz_4^n \rtimes \bbz_2^{\cal{C}} \to  \bbz_2^n  \times \bbz_2^{\cal{C}} \to 1.
\label{ex3}
\ee
This can be done by summing over sectors twisted by  the $\bbz_2^n $ subgroup both in space and time. The gauging is equivalent to shifting the theory $S[R]$ to $S[R/2]$ \cite{ji2020top}. As a result, if we fix $R$, gauging $\bbz_2^n$ subgroup of $\bbz_4^n$ maps $V_{n, w} \to V_{2n, w/2}$. The gauged theory has the symmetry $\bbz_2^n \times \bbz_2^w \times \bbz_2^{\cal{C}}$ where we have identified $\hat{\bbz}_2^n$ with $\bbz_2^w$. Further gauging $\bbz_2^n$ yields $V_{2n, w/2} \to V_{4n, w/4}$. In other words, gauging $\bbz_4^n$  yields $\bbz_4^w \equiv\hat{\bbz}_4^n$, the dual symmetry of the original theory, $V_{n, w} \to V_{4n, w/4}$. Thus the clock spin operator on a lattice  can be identified with $V_{1/2, 0}$, and the disorder operator  can be identified with the nonlocal $V_{0, 1/2}$ (up to higher order corrections). After gauging $\bbz_4^n$, the roles of the order operator and the disorder operator are exchanged. However, if we only gauge the $\bbz_2^n$ subgroup,  the order operator and the disorder operator of the quotient $\bbz_2^n$ are $V_{1, 0}$ and $V_{0, 1/2}$, respectively.  Similarly, the order operator and the disorder operator of $\bbz_2^w$ are $V_{0, 1}$ and $V_{1/2, 0}$, respectively. 

We should compare the mapping to the correspondences in Table \ref{table1}. After gauging, the order operator $V_{1/2, 0}$ of the original $\bbz_4^n$ becomes the disorder operator of the dual  $\bbz_2^w$ while the disorder operator $V_{0, 1/2}$ remains the disorder operator of the quotient $\bbz_2^n$. 
There is a mixed anomaly between $\bbz_2^n  $ and $\bbz_2^w$, meaning that gauging one symmetry breaks the other. An implication of this anomaly is that two copies of the disorder operator $V_{0, 1/2}$ of $\bbz_2^n$ are exactly the order operator $V_{0, 1}$ of $\bbz_2^w$ and vice versa. Thus, the disorder operator $V_{0, 1/2}$ of $\bbz_2^n$  can be viewed as a (fractional) order operator $V_{0, 1}$ of $\bbz_2^w$ and vice versa. These relations are exactly what we observed in section \ref{subsec_gauge1}.
 
The discussion above can be found extensively in the literature \cite{karch2019web, ji2020top, gaiotto2021orbifold, thorngren2019fusion}. What we did is reverse the story. Gauging the $\bbz_2^n$ gives back the original symmetry.  We can also gauge $\bbz_2^{\cal{C}}$ instead. This story, together with the Kramers-Wannier duality and its generalizations (non-invertible symmetries), as well as its corresponding electromagnetic duality in the $(2+1)D$ bulk, is interesting itself, but we refer the reader to Ref. \cite{karch2019web, ji2020top, thorngren2019fusion}. In addition, the connection to categorical symmetries can be found in Ref. \cite{chatterjee2023symmetry, chatterjee2022holographic}. We would like to point out that gauging finite subgroups of a symmetry in fermionic theories such as the free Dirac theory \cite{karch2019web, ji2020top, gaiotto2021orbifold} can be viewed as an extension of the discussion in this work.

\section{Clock model in  $(2+1)$D}
\label{2D}
 Now we extend the discussion in the previous section to $(2+1)$D. For concreteness, we start again with the 4-state clock model on a 2D square lattice with periodic boundary conditions (Fig.\ref{lattice}): 
  \be 
H  =- \sum_{{\bd{ r}}, {\bd{ r}}' \in \partial {e_{{\bd{ r}}, {\bd{ r}'}}}} (C^{\dagger}_{\bd{ r}} C_{{\bd{ r}}'}+  h.c.)  - g \sum_{\bd{ r}} (S_{\bd{ r}} +h.c.)   ,
\label{clock2}
\ee
where $e_{{\bd{ r}}, {\bd{ r}'}}$ is a link connecting two sites $\bd{r}$ and $\bd{r}'$. $C_j$ and $S_j$ satisfy the same algebra in Eq.(\ref{weyl}). Then the model has a $\bbz_4$ symmetry $U =\prod_{\bd{ r}} S_{\bd{ r}}$ and has a second-order phase transition by tuning $g$ because similar to its 1D version it is equivalent to two decoupled Ising models. Numerical calculation shows that $g_c=3.044$ \cite{Blote2002cluster,  Zhao2021}. The order parameter and the disorder parameter in the clock model are, respectively, 
\be O =\lim_{|{\bd{ r}}-{\bd{ r}'}|\gg 1} \langle C^{\dagger}_{\bd{ r}} C_{\bd{ r}'}\rangle \ee and  \be D =\lim_{|M| \gg 1} \langle \prod_{\bd{r} \in M}  S_{\bd{ r}}\rangle, \ee
where $M$ is a connected area and $|M|$ is its area. In the symmetry breaking phase, $O$ is a constant and $D$ has, in the leading order, an area law $D \sim \exp[- \alpha |M|]$, where  $\alpha$ is a constant. In the disordered phase, $O$ vanishes and, in the leading order,  $D\sim \exp[- \beta |\partial M|]$, where $|\partial M|$ is the perimeter of the boundary $\partial M$ and $\beta$ is a constant.
For a recent study in the Ising model along this direction, see Ref.\cite{Zhao2021}. 

\subsection{Gauging $\bbz_2$}
\label{subsec_gauge2}
\begin{figure}[t]
\includegraphics[width=0.3\textwidth]{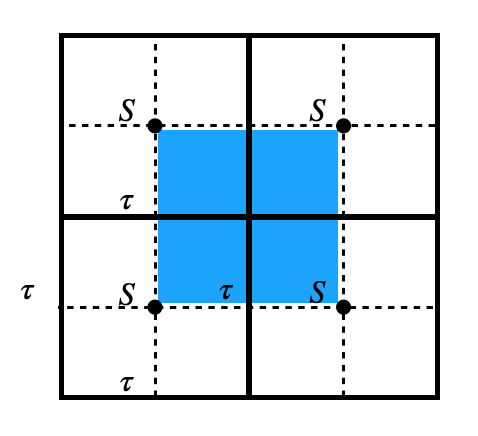}
\centering
\caption{Schematic of the original square lattice $\Sigma$ (dashed) and its dual lattice $\Sigma^*$ (solid) . Original clock spins sit on each site of $\Sigma$ and the gauge spins sit on the links of $\Sigma$ and $\Sigma^*$. The condition $\prod_{e\in \partial p} \tau_e^z =1$ is imposed on each plaquette of $\Sigma$, e.g. highlighted in blue. }
\label{lattice}
\end{figure}

Now we gauge the clock model by placing $\tau^z_e$ on links:
 \be 
H  = - \sum_{e} (C^{\dagger}_{\bd{ r}}\tau^z_{e_{{\bd{ r}}, {\bd{ r}'}}} C_{{\bd{ r}}'}+  h.c.)  - g \sum_{\bd{ r}} (S_{\bd{ r}} +h.c.)  - \lambda \sum_p \left( \prod_{e\in \partial p} \tau_e^z -1\right)
\label{gauged2}
\ee
 with the constraints 
  \be 
 \prod_{\bd{r} \in \partial e}  \tau_e^x = S_{\bd{r}}^2.
 \label{constraints2}
 \ee
 Here, $p$ denotes any plaquette of the original lattice (Fig.\ref{lattice}).
 We take the large $\lambda$ limit to restrict the Hilbert space to the zero flux sector, i.e. $\prod_{e\in \partial p} \tau_e^z =1$, so that the gauge field is flat. The gauged theory has a 0-form $\bbz_2$ symmetry group  generated by \be U_1 =\prod_{\bd{ r}} S_{\bd{ r}}\ee
and a 1-form $\bbz_2$ symmetry generated by the Wilson loop
\be
U_2 = \prod_{e \in \gamma} \tau_e^z,
\ee
where $\gamma$ is a closed loop on the original lattice \footnote{The existence of a 0-form symmetry and a 1-form symmetry in the theory may remind us of 
2-groups (or obstructions to symmetry fractionalization) \cite{kapustin2014anomalous, kapustin2014anomalies, Barkeshli2019symmetry, benini2019on} which correspond to the second case in Ref.\cite{tachikawa2020}.  However, $G$ is anomaly-free in our case and it corresponds to the first case in Ref.\cite{tachikawa2020}.}. The large $\lambda$ limit renders the 1-form symmetry exact because then $U_2$ only depends on the topology of $\gamma$. $U_2$ can take values of $\pm 1$ if $\gamma$ is noncontractible on the torus.  Its spontaneous symmetry breaking implies deconfinement \cite{gaiotto2015generalized, mcgreevy2023generalized}. In particular, its spontaneous breaking implies a degeneracy depending on the topology of the lattice. 

Once again in the gauged theory, $O$ becomes
\be O =\lim_{|{\bd{ r}}-{\bd{ r}'}|\gg 1} \langle C^{\dagger}_{\bd{ r}} \prod_{e\in \gamma, \bd{ r}, \bd{ r}' \in \partial \gamma } \tau_e^z C_{\bd{ r}'} \rangle. \ee
The edge operators create charges under $U_1$.
Also, using the constraints in Eq.(\ref{constraints2}), we obtain 
\be
 \prod_{\bd{r} \in M}  S^2_{\bd{ r}} =  \prod_{e \perp \partial M}  \tau_e^x,
 \ee
where $\partial M$ is viewed as a loop on the dual lattice, and $e \perp \partial M$ denotes all links pierced by the loop. This equality indicates that  $\prod_{\bd{r} \in M}  S_{\bd{ r}} $ as a disorder operator is locally fractionally charged under $U_2 = \prod_{e \in \gamma} \tau_e^z$ if $\gamma$ and $\partial M$ intersect. However, as long as $M$ is contractible, it remains overall neutral: $U_2$ measures the magnetic flux enclosed by $\gamma$.

\subsection{Eliminating the Gauss-law condition}
\label{map2}
Similar to the $(1+1)$D case, we can eliminate the Gauss-law condition in Eq.(\ref{constraints2}). Then
\be S_{\bd{r}} \to \left(\frac{1+ \prod_{{\bd{r}} \in \partial e} \tau_{e}^x}{2}\right) \sigma_{\bd{r}}^z + \left(\frac{1-\prod_{{\bd{r}} \in \partial e} \tau_{e}^x }{2}\right) i \sigma_{\bd{r}}^z,
\ee
and
\be 
 C_{\bd{r}}^{\dagger} \tau^z_{e_{{\bd{ r}}, {\bd{ r}'}}}       C_{{\bd{r}}'}    \to    K_{\bd{r}}^{\dagger} \tau_{e_{{\bd{ r}}, {\bd{ r}'}}}^z K_{\bd{r}'},
\ee
where 
\be 
K_{\bd{r}} =  \left(\frac{1+ \prod_{{\bd{r}} \in \partial e} \tau_{e}^x }{2}\right) \sigma_{\bd{r}}^x + \left(\frac{1-\prod_{{\bd{r}} \in \partial e} \tau_{e}^x }{2}\right).  
\label{kk1}
\ee  
We simply replaced $\tau_{j-1/2}^x \tau_{j+1/2}^x$ in the $(1+1)$D case with  $\prod_{\bd{r} \in \partial e}  \tau_e^x$.
Under this map, 
\be 
U_1 =\prod_{\bd{ r}} S_{\bd{r}} \to  U_a = \prod_{\bd{ r}} \sigma_{\bd{r}}^z \ i^{\frac{ 1- \prod_{{\bd{r}} \in \partial e} \tau_{e}^x}{2}}      , \quad U_2 =\prod_{{e_{{\bd{ r}}, {\bd{ r}'}}} \in \gamma}  C_{\bd{r}}^{\dagger} \tau_{e_{{\bd{ r}}, {\bd{ r}'}}}^z       C_{{\bd{r}}'}  \to   U_b = \prod_{e \in \gamma} \tau_e^z.
\ee
It can be checked again that $U_a$ still generates a $\bbz_2$ symmetry because flipping a spin $\tau_{e}^x$  necessarily changes $\sum_{\bd{r}}\prod_{{\bd{r}} \in \partial e} \tau_{e}^x$ by a multiple of 4. The reduced Hamiltonian is 
 \begin{align}
H  =  - \sum_{e} 
(K_{\bd{ r}}^{\dagger} \tau_{e_{{\bd{ r}}, {\bd{ r}'}}}^z K_{\bd{ r}' } +K_{\bd{ r}'}^{\dagger} \tau_{e_{{\bd{ r}}, {\bd{ r}'}}}^z K_{\bd{ r} } )  
  - g \sum_{\bd{ r}} \left(\frac{1+ \prod_{{\bd{r}} \in \partial e} \tau_{e}^x}{2}\right) \sigma_{\bd{r}}^z
- \lambda \sum_p \left( \prod_{e\in \partial p} \tau_e^z -1\right) .
\end{align} 
Similar to its   $(1+1)$D analogue, the phase transition at the critical point is inherited from its ungauged counterpart as in Fig.\ref{phases}. The difference is that now the critical point is between a  0-form symmetry and a 1-form symmetry. In other words, it is an unconventional critical point between an ordinary symmetry breaking phase and a topologically ordered phase (enriched by the 0-form symmetry) .
The scaling laws in the correlation functions of order operators and disorder operators are also directly derivable from those in the original clock model.
 
Note that in the reduced Hamiltonian, the zero flux condition, i.e. $\lambda\to \infty$,  now becomes effectively the new gauge constraint on $\tau^x_e$ on the dual lattice. It not only guarantees that the total Hilbert space dimensions of the original clock model and of the reduced Hamiltonian match but also that the 1-form symmetry $U_2 = \prod_{e \in \gamma} \tau_e^z$ remains exact. $U_2$ is the 't Hooft loop operator with respect to the ``new" gauge field $\tau^x_e$ and its spontaneous symmetry breaking corresponds to a topological order.  The gauged model can be viewed as a $\bbz_2$ gauge theory coupled to  dynamical $\bbz_2$ magnetic fluxes decorated with $\bbz_2$ 0-form charges.

\subsection{Anomaly action and the SPT bulk}
\label{subsec_anomaly}
We have constructed a concrete lattice model such that the 0-form symmetry and the 1-form symmetry it has have a mixed anomaly. The theory can be interpreted as the boundary theory of a SPT phase protected by these two symmetries in the $(3+1)$D bulk $Y$. The anomaly action is given by \cite{tachikawa2020}
\be 
 \int_{Y}  \hat{a} \cup e(g) \in U(1),
\label{anomaly1}
\ee
where $\hat{a} \in H^2(Y, \hat{\Gamma})$, $g \in Z^1(Y, A)$ is the background gauge field of $A$ extended from $X =\partial Y$ to the bulk $Y$, $e(g) \equiv g^*(e) \in H^2(Y, \Gamma) $ is the element defined by the extension class $e \in H^2(A, \Gamma) \cong H^2(BA, \Gamma)$ of the sequence (\ref{seq1}) pulled back by the map $Y \to BA$ corresponding to $g$. Here, $BA$ is the classifying space of $A$. $a \in C^1(Y, \Gamma)$ is the background field of $\Gamma$,  and satisfies $\delta a  = e(g)$. Original discussions can be found in Ref.\cite{gaiotto2017theta, kapustin2017higher}.

SPT phases protected by higher form symmetries have been discussed in many references \cite{gaiotto2015generalized,  Yoshida2016top, kapustin2017higher, Tsui2020lattice}.
Note that similar anomaly actions were also discussed in Ref.\cite{jian2021physics} in the continuum version even though the connection to gauging a subgroup was not included. Ref.\cite{jian2021physics} used the analog of the ``flux attachment" construction of bosonic SPT phases \cite{Senthil2013integer} and constructed a $(3+1)$D SPT phase protected by a 0-form symmetry $U(1)$ and a 1-form symmetry $U(1)^{(1)}$ by decorating the Dirac monopole with a zero-dimensional bosonic SPT state with $U(1)$ symmetry and then condensing the decorated Dirac monopoles. Equivalently, the SPT state can be obtained by condensing the vortex lines of the $U(1)$ symmetry decorated with a one-dimensional bosonic SPT with $U(1)^{(1)}$ symmetry.  The bulk SPT protected by $U(1)^{(1)} \times U(1)$ is then described by the anomaly/response action
\be 
S  = \int_Y \frac{i k}{2\pi} B\wedge dA,
\label{anomaly2}
\ee
where $B$ and $A$ are the 2-form and 1-form background gauge fields of $U(1)^{(1)} $ and $U(1)$, respectively. The SPT phase protected by discrete symmetries $\bbz_q^{(1)}$ and $\bbz_n$ can be obtained as a descendant by introducing $q$-fold dynamical electric charges and replacing the zero-dimensional bosonic SPT state protected by $U(1)$ symmetry with one protected by the $\bbz_n$ symmetry.  If $q =n=2$ and $k=1$, we can replace the continuous gauge fields with discrete gauge fields $B \to \pi b$ and $A \to \pi \tilde{g}$ (and also $\wedge \to \cup$ and $d\to \delta$) \cite{gaiotto2017theta}. Here, $\tilde{g}$ is viewed as an integer-valued cochain that is closed modulo 2. The anomaly in Eq.(\ref{anomaly2}) becomes
$S  = i (\pi/2) \int_Y b \cup \delta \tilde{g}$. In terms of the central extension (\ref{seq1}), $e(g)=\beta(g)=g \cup g$ where $\beta$ is the connecting homomorphism $H^1(Y, \bbz_2) \to H^2(Y, \bbz_2)$, i.e. $\beta(g) =\delta \tilde{g}/2 \mod 2$ (see Appendix \ref{append2}).  Thus, the anomaly in Eq.(\ref{anomaly2}) reduces to $S  = i \pi \int_Y b \cup e(g)$ which is the same as  that in Eq.(\ref{anomaly1}) once we identify $b$ with $\hat{a}$ and decode the Pontryagin duality \cite{Kapustin2014}. The boundary theory is identified with a $\bbz_2$ gauge theory with no dynamical $\bbz_2$ electric charges and with  magnetic fluxes charged under the $\bbz_2$ 0-form symmetry, the same as in the gauged clock model. Here, we have only provided a physical description of the SPT bulk in the continuum language, and 
it would be interesting to write down an explicit lattice realization of the SPT bulk in the future \cite{Yoshida2016top}. What we have shown is that the boundary theory can have an unconventional critical point.

Note that even though we have focused on the $\bbz_2^{(1)} \times \bbz_2$ case, the response theory in Eq.(\ref{anomaly2}) of the $(3+1)$D SPT phase descends to more general cases. Indeed, in Ref.\cite{jian2021physics}, it was argued that the SPT phase protected by $\bbz_q^{(1)}$ and $\bbz_n$ is classified by $\bbz_{\text{gcd}(q, n)}$. Unsurprisingly, this is the same as the extension class $H^2(\bbz_q, \bbz_n) =\bbz_{\text{gcd}(q, n)}$. Similarly,  it was argued that the SPT phase protected by $\bbz_q^{(1)}$ and $U(1)$ as in Eq.(\ref{anomaly2}) is classified by $\bbz_q$, which is again consistent with $
H^2_B(U(1), \bbz_q) = H^2(BU(1), \bbz_q) =H^2(CP^{\infty}, \bbz_q) = \bbz_q.
$ Here, in order to consider continuous $U(1)$, we used the Borel group cohomology \cite{chen2013}. In the second equality, we used the fact $BU(1) =CP^{\infty}$, the infinite complex projective space 
\cite{stasheff1978continuous, toda1987cohomology}. Therefore, the idea of gauging a finite subgroup is not restricted to finite $G$.

\subsection{Similarity to the $SU(2)$ mixed gauge theory}
\label{subsec_mixed}

As in the section \ref{d8}, the total symmetry of the original 4-state clock model contains $D_8$. The anomaly structure is very similar to that in the $SU(2)$ mixed gauge theory discussed in Ref.\cite{gaiotto2017theta}. There, they reduced the $SU(2)$ gauge theory with $\theta =\pi$ at finite temperature from $(3+1)$D to $(2+1)$D by compactifying the imaginary time direction. The reduced theory in $(2+1)$D has a 0-form symmetry associated with the CP symmetry, a 0-form symmetry and a 1-form symmetry from the original 1-form center symmetry. After gauging the 1-form symmetry, the mixed gauge theory has an anomaly-free symmetry $D_8$, similar to the ungauged clock model. Thus, the story we had can be viewed as reversing the gauging process. Indeed, gauging the center symmetry $\bbz_2 \subset D_8$, we obtain the the reduced theory with $ \bbz_2^{(1)} \times \bbz_2^{(0)} \times \bbz_2^{(0)}   $ with the anomaly $i\pi \int_Y    B \cup e(g) $ where $e(g) =g_1\cup g_2$ and $g = (g_1, g_2) \in Z^1(Y, \bbz_2^{(0)} \times \bbz_2^{(0)})$.

There is a subtlety here. As we mentioned in section \ref{d8},  there are three order 2 elements in $\bbz_2^a \times \bbz_2^{\cal{C}'}$ and any two of them can be viewed as the two generators.  This fact leads to three inequivalent extensions associated with the sequence (\ref{seq2}), depending on which two elements in $\bbz_2^a \times \bbz_2^c$ the generators $S$ and $\cal{C}$ are projected onto (see Appendix \ref{append2}). If $e(g) =g_1\cup g_2$ when $g = (g_1, g_2)$, we can choose a different $g' = (g_1, g_1+g_2)$, then $e(g') =g_1\cup g_1   +g_1\cup g_2$. It reduces to the discussion in the previous subsection if $g_2 =0$.  

In Ref.\cite{gaiotto2017theta}, they discussed different possible cases with different number of vacua. In our case, there are 4 vacua across the continuous transition. Similar to the discussion in Ref.\cite{gaiotto2017theta},  the properties of the domain walls between these vacua should be constrained by the anomaly. In addition, 
one can reduce the $SU(2)$ gauge theory to a $\bbz_2$ gauge theory by introducing Higgs fields while preserving the $\bbz_2$ 1-form symmetry \cite{wang2020gauge, bi2020landau}. Then the model in Eq.(\ref{gauged2}) may be regarded as an effective model where an unconventional transition can take place. What we emphasized in this work is that the transition can be continuous.

\section{General case}
\label{general}
The analysis in previous sections can obviously be generalized to higher dimensions. In this section, we briefly discuss the general case in an arbitary number of dimensions. We will focus primarily on $G =\bbz_n$. Generalizations to other discrete and continuous symmetries, e.g. $U(1)$, may need some modifications of language.  We will try to argue that, in general, if a theory $\cal{T}$ has a continuous critical point where $G$, for simplicity, is completely spontaneously broken, the gauged theory ${\cal{T}}/\Gamma$ has an unconventional critical point associated with $A$ and $ \hat{\Gamma}^{(d-1)}$ if $G$ is a nontrivial extension of $A$ by $\Gamma$ \footnote{It was argued in Ref.\cite{bhardwaj2022universal, bartsch2022non} that in general gauging a finite subgroup yields a dual symmetry that has a $(d-2)$-category structure.}. 

Suppose that $\cal{T}$ is defined on a $d$-dimensional  hypercubic lattice $\Sigma$ with periodic boundary conditions. The action of the symmetry $G=\bbz_n$ is on-site and implemented on $\cal{T}$ by $R=\prod_{\bd{r} \in \Sigma} S_{\bd{r}}$. For example, in the clock model, $S_{\bd{r}}$ is  the shift operator.  $O_{\bd{r}}$ is a local Hermitian operator
that transforms faithfully under $G$, i.e.  $R_g O_{\bd{r}} R_g^{-1} = g O_{\bd{r}} $ where $g \in G$ with $G$ embedded into $U(1)$.  An onsite representation of $G$ also means that $G$ is nonanomalous. We take the volume (total number of sites) of the lattice $|\Sigma|$ to be large.   Across the continuous critical point, $G$ is either preserved or fully broken.  The two-point correlation function $\lim_{|\bd{r}-\bd{r}'|\gg 1}\langle O_{\bd{r}}^{\dagger} O_{\bd{r}'} \rangle$, neutral under $R$, serves as an indicator of the ordinary phase transition. In the ordered phase, it approaches a constant while in the disordered phase, it decays exponentially to 0. In the ordered phase, for a large $M$, the operator $\prod_{\bd{r} \in M} S_{\bd{r}}$ creates domain walls and can be regarded as the disorder operator of the symmetry-breaking phase. In general, $\langle \prod_{\bd{r} \in M} S_{\bd{r}} \rangle$ can have an area law or a perimeter law in the leading order.  If  $\langle \prod_{\bd{r} \in M} S_{\bd{r}} \rangle\sim \exp[- \alpha |M|]$, where $|M|$ is the area of $M$,  then the phase is ordered. If $\langle \prod_{\bd{r} \in M} S_{\bd{r}} \rangle\sim \exp[- \beta|\tilde{\gamma}|]$, where $\tilde{\gamma} = \partial M$, then the domain walls proliferate and the symmetry is preserved.

\begin{figure}[t]
\includegraphics[width=0.6\textwidth]{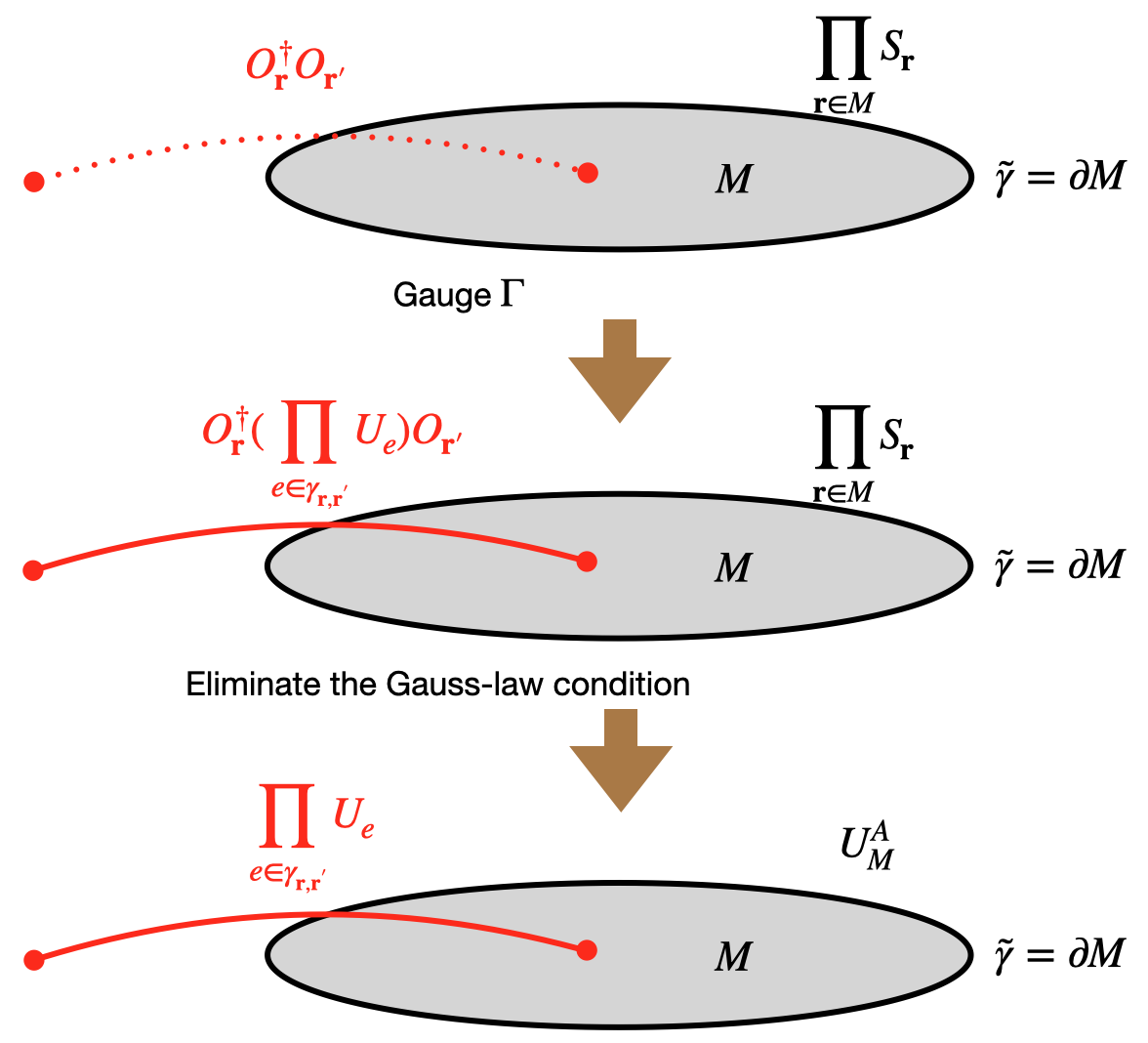}
\centering
\caption{Correspondences of order/disorder operators before and after gauging a finite sbugroup $\Gamma$ and eliminating the Gauss-law condition.  Upper: $\prod_{\bd{r} \in M} S_{\bd{r}}$ is the disorder operator  of $G$ and $O_{\bd{r}}^{\dagger} O_{\bd{r}'} $ (red points with dashed line) is used in the order parameter of $G$. Middle/Lower: $\prod_{\bd{r} \in M} S_{\bd{r}}$ or $U_M^A$ is the disorder operator  of $A$ and $O_{\bd{r}}^{\dagger} 
( \prod_{e \in \gamma_{\bd{r}, \bd{r}'}} U_e )  O_{\bd{r}'}$ or $\prod_{e \in \gamma_{\bd{r}, \bd{r}'}} U_e$ (red points with solid line) the disorder operator of $\hat{\Gamma}^{(d-1)}$.}
\label{disorders}
\end{figure}
 
Now we gauge the $\Gamma$ subgroup of $G$ to obtain ${\cal{T}}/\Gamma$ such that the symmetries  $G/\Gamma$ and $ \hat{\Gamma}^{(d-1)}$ have a mixed anomaly (Fig.\ref{disorders}). Coupling a theory $\cal{T}$ to flat background gauge fields amounts to twisting the boundaries. Gauging $\Gamma$ amounts to summing over twisted sectors with some weights. This process is analogous to orbifolding in $2$D. In the gauged theory, $\prod_{\bd{r} \in \Sigma} S_{\bd{r}}$ still generates $A  =G/\Gamma$. The  Wilson loop operators for the gauged symmetry are the generators of $\hat{\Gamma}^{(d-1)}$ and the charged objects are the twisted sector operators \cite{gaiotto2015generalized}.  Denote the gauge field as $U_e$, e.g., $U_e = \sigma^z_e$ in the $\bbz_2$ gauge field. Then the generators of the dual $\hat{\Gamma}^{(d-1)}$ symmetry are
$\prod_{\gamma} U_e$, where $\gamma$ is an arbitrary closed loop. The charged objects are $\prod_{\bd{r} \in  M} S_{\bd{r}}$ with a $(d-1)$-dimensional boundary. Then $
  O_{\bd{r}}^{\dagger} 
( \prod_{e \in \gamma_{\bd{r}, \bd{r}'}} U_e )  O_{\bd{r}'} $, where $\gamma_{\bd{r}, \bd{r}'}$ is a shortest path connecting $\bd{r}$ to $\bd{r}'$, can be taken as the disorder operator of $\hat{\Gamma}^{(d-1)}$ and   $  \prod_{\bd{r} \in M} S_{\bd{r}}  $ can be viewed as the disorder operator of $A$.  These are called ``patched operators" in Ref.\cite{ji2020}. If there is an intersection between $\gamma$ and $\tilde{\gamma}= \partial M$, then these two operators do not commute. In other words, there is a ``mutual statistics" between them \cite{ji2020}.  As in section \ref{map1} and \ref{map2}, we may further eliminate the Gauss-law condition by factoring the Hilbert space on each site (Fig.\ref{disorders}). In this case, due to the mixed anomaly, the inherited disorder operator $U_M^A$ of  $A$ and $\prod_{e \in \gamma_{\bd{r}, \bd{r}'}} U_e$ of $\hat{\Gamma}^{(d-1)}$ cannot be both onsite.  Condensation of disordered operators of one symmetry leads to SSB of the other symmetry. We obtain the scenario in Fig.\ref{phases}.

Since gauging a finite field amounts to coupling the theory to some discrete topological degrees of freedom, it is expected that scaling laws of the order/disorder parameters are unaffected \cite{Kapustin2014}. In fact, for each flat background gauge field $U_{\bd{r}}$, the  scaling law of  $
\lim_{|\gamma_{\bd{r}, \bd{r}'}|\gg 1} \langle O_{\bd{r}}^{\dagger} 
( \prod_{\gamma_{\bd{r}, \bd{r}'}} U_{\bd{r}} )  O_{\bd{r}'} \rangle$ and  of $
\lim_{|\gamma_{\bd{r}, \bd{r}'}|\gg 1} \langle O_{\bd{r}}^{\dagger} O_{\bd{r}'} \rangle$ should match because we can choose the gauge $U_{\bd{r}} =\text{const.}$ locally due to the flatness of the connection \cite{Kapustin2014}. Summing over different flat background gauge fields/twisted boundary conditions does not change it. 
Also, since we assumed that the phase transition of $\cal{T}$ is ordinary, the scaling laws of $\lim_{|\gamma_{\bd{r}, \bd{r}'}|\gg 1} \langle O_{\bd{r}}^{\dagger} O_{\bd{r}'} \rangle$ and $\lim_{|M|\gg 1} \langle \prod_{\bd{r} \in M} S_{\bd{r}} \rangle$ do not depend on boundary conditions. Consequently, the scaling laws remain identical in $\cal{T}$ and in ${\cal{T}}/\Gamma$. The correspondences across the continuous critical point are summarized in Table \ref{table1}.

The relations between order operators and disorder operators remind us of the categorical symmetry discussed in Ref.\cite{ji2020}. The difference is that in their work, they focused on  gauging the entire symmetry group. Restricted to the symmetric sub-Hilbert space, patched operators associated with the symmetry and its dual have nontrivial relations, revealing a nontrivial categorical symmetry. The non-tensor-product-like Hilbert space can be viewed as a system with non-invertible gravitationaly anomaly and as the boundary of a topologically ordered phase and the categorical symmetry is directly related to the conservation of gauge charges and of gauge fluxes. One consequence is that it is possible that around the critical point where the categorical symmetry is preserved, either the symmetry or its dual is spontaneously broken but not both. However, in our discussion, we did not restrict the Hilbert space to the symmetric sector. As a result, the criticality occurs at the boundary of a SPT phase instead of a topologically ordered phase.    Nevertheless, we can also discuss the categorical symmetry of the system after gauging a finite subgroup. It was suggested that the categorical symmetry is unaffected by the gauging process \cite{chatterjee2023symmetry, chatterjee2022holographic}.

If we want to consider  $G=U(1)$, e.g. for the $XY$ model or the quantum rotor model, then the disorder operator is the vortex operator which is charged under the dual symmetry \cite{gaiotto2015generalized}. We expect that similar analysis as in this section should follow. The low energy effective theory is described by a complex scalar field as in superfluids \cite{Dela2020super}. Similar to the discussion in section \ref{sec_boson}, we may analyze the consequence of gauging a finite subgroup. 

\section{Discussion}  
The idea presented in this work is very general and has many interesting implications. 
Many unconventional critical points can potentially be constructed from ordinary critical points. 
Constructing DQCPs in $(1+1)$D
\cite{jiang2019ising, roberts2019deconfined, huang2019emergent, mudry2019quantum, roberts2021one,  zhang2023, lee2022landau} is an immediate application. These critical points, as a result of 't Hooft anomalies, can be viewed as boundary critical points of SPT phases. They should be compared with the phase transitions allowed by  the Lieb-Schultz-Mattis theorem \cite{lieb1961, hastings2004, oshikawa2000} and its higher-form generalization \cite{Kobayashi2019} where the (non-onsite) translational symmetry and an internal symmetry have a mixed anomaly when they act on the low-energy degrees of freedom\cite{cho2017anomaly, Metlitski2018}.  We can also interpret the mixed anomaly from gauging a subgroup as an emergent anomaly where the gauge constraints are imposed energetically \cite{wang2018, Metlitski2018, thorngren2021}.

We would like to emphasize that the argument does not depend on the detailed lattice realization but only on the general symmetry. For example, instead of starting with the clock model in Eq.(\ref{clock1}), we can start with the quantum Potts model \cite{cobanera2011bond}
and the analysis remains largely unaffected as long as the the $\bbz_4$ symmetry is spontaneously broken. 
Since the general argument depends only on the symmetry of the Hamiltonian, the ungauged Hamiltonian can be complicated. For example, we discussed in section \ref{subsec_mixed} that the gauged 4-state clock model has the same mixed anomaly as the compactified $SU(2)$ gauge theory with $\theta = \pi$ in 3D, and the original 4-state clock model corresponds to the mixed gauge theory. Nevertheless, the transition points before and after gauging the finite subgroup are still directly related. A second order critical point remains after gauging. Moreover, since topological ordered phases are robust to weak explicit breaking of the higher-form symmetries \cite{gaiotto2015generalized, mcgreevy2023generalized}, we may expect such unconventional critical points to be stable.

 We can also reverse the argument.
Suppose we have a SPT phase in $(d+1)$D protected by $A \times \hat{\Gamma}^{(d-2)}$ and we have obtained the effective boundary theory where there is a mixed anomaly between $A$ and $\hat{\Gamma}^{(d-2)}$. We can gauge the $\hat{\Gamma}^{(d-2)}$ symmetry to obtain a new 0-form symmetry $G$, a group obtained by extending $A$ by $\Gamma$. If the new symmetry is nonanomalous, the critical point on the SPT boundary can be mapped to an ordinary critical point.

There are many directions to generalize the analysis. In terms of concrete lattice models, a generalization to the $q$-state clock model with $q >4$ is possible, but the next nontrivial case is $q =8$, since the extension class is trivial for $5\le q \le 7$. There is an  emergent $U(1)$ symmetry around the critical point
\cite{lou2007emergence, shao2020monte, patil2021unconventional}. It would also be helpful to construct microscopic models of the SPT bulk instead of the continuum analysis as in section \ref{subsec_anomaly} such that these clock models are realized as the boundaries. 

 Also, it is useful to construct concrete lattice models for more general discrete or continuous $G$. For example, as we mentioned in section \ref{d8}, we may couple the 4-state clock model to the Ising model such that the total symmetry group is $D_8$ and it is completely broken via a continuous transition point by tuning the parameters. Since $D_8$ is the symmetry of the square, it is natural to construct models on a square lattice such that the $D_8$ symmetry is preserved. However, a continuous transition that breaks $D_8$ directly and completely often needs fine-tuning \cite{Takahashi2020valence}. In addition, we can consider the $U(1)$ case associated with the $XY$ model or the quantum rotor model. The theory of the SPT bulk should be obtained by properly generalizing (and discretizing) the analysis in Ref.\cite{jian2021physics}.
 
Furthermore, a detailed analysis about the correspondence between the Landau-Ginzburg-Wilson theory around the ordinary critical point before gauging and the theory that describes the corresponding unconventional critical point can be very useful. If the critical point is described by a CFT, the gauged critical point is described by the gauged CFT. The corresponding (possibly isomorphic) transformation on the categorical symmetry 
\cite{ji2020, chatterjee2023symmetry, chatterjee2022holographic} and its interplay with dualities deserve further explorations. 

We can also consider generalized symmetry $G$ \cite{mcgreevy2023generalized, schafer2023ictp} or non-Abelian $\Gamma$ \cite{bhardwaj2018finite, bhardwaj2022universal, bartsch2022non}, and generalizations to fermionic models. We leave these questions for future work.

Overall, we believe a systematic study in these directions will uncover more  unconventional critical points both in condensed matter physics and in high energy physics.

\acknowledgments
The author thanks Michael Levin for discussion, Yi-Zhuang You and Meng Zeng for comments, and Ivar Martin and Aashish Clerk for support.

\appendix
\section{Trivial case: gauging $\bbz_2$ of the 6-state clock model}
\label{6clock}
Since $\bbz_6 \cong \bbz_2 \times \bbz_3$, the extension of $\bbz_3$ by $\bbz_2$ is trivial. More generally, the extension of $\bbz_q$ by $\bbz_n$  is classified by $H^2(\bbz_q, \bbz_n) = \bbz_{\text{gcd(q, n)}}$, where $\text{gcd}(q, n)$ is the greatest common divisor of $q$ and $n$. If $q$ and $n$ are co-prime, the extension is trivial. Let us see how the triviality displays itself upon gauging $\bbz_2$ of the 6-state clock model: 
\be 
H  = - \sum_j (C^{\dagger}_j\tau^z_{j+1/2} C_{j+1}+  C^{\dagger}_{j+1} \tau^z_{j+1/2} C_j)  - g \sum_j (S_j +S_j^{\dagger}),
\label{gauged3}
\ee
 with the condition
 \be 
 \tau^x_{j-1/2} \tau^x_{j+1/2} = S_j^3.
 \label{constraints3}
 \ee
 The algebra of $C_j$ and $S_j$ is 
 \be 
C_j^6  = S_j^6 = 1, \quad C_j S_j  = \omega_6 S_j C_j,
\label{weyl3}
\ee 
where $\omega_6 = \exp(i2\pi/6 )$.  The simplest way to see two untangled symmetries is to use
  \be 
U_1  = \prod_j S_j^2, \quad U_2 = \prod_j \tau^z_{j+1/2}.
 \ee
No constraints are needed to show that $U_1^3 =1$. We may also use 
 \be 
U_1  = \prod_j S_j, \quad U_2 = \prod_j \tau^z_{j+1/2}.
 \ee
 The constraints in Eq.(\ref{constraints3}) then imply $U_1^3 =1$ when the periodic boundary condition is used.
Superficially, we may think that the $\bbz_2$ and $\bbz_3$ symmetries have a mixed anomaly. However, we can eliminate the Gauss-law condition as in section \ref{map1} by mapping a 6-state clock to one Ising spin and a 3-state clock spin where they are both diagonalized: 
 $|S_j \rangle$: $|S_j \rangle \to |\tau^x_{j-1/2}\tau^x_{j+1/2}, S^{(3)}_j\rangle$. Here, $S^{(3)}_j$ is the shift operator of the new 3-state clock model.  It is easy to see that the symmetry generators are onsite: 
$ U_a = \prod_j S^{(3)}_j$ and $ U_b = \prod_j \tau_{j+1/2}^z$.

\section{Central extensions of $ \bbz_2 \times \bbz_2$ by $\bbz_2$}
\label{append2}
Given an extension of $ \bbz_2 \times \bbz_2$ by $\bbz_2$,
\be 
1 \to \bbz_2 \to G \to  \bbz_2 \times \bbz_2 \to 1,
\ee
we can find a cocycle representation of  $e(g) \equiv g^*(e) \in H^2(Y, \bbz_2) $ defined by the central extension class $e \in H^2(A, \bbz_2) = H^2(BA, \bbz_2)$ pulled back by the map $Y \to BA$ corresponding to the cocycle $g \in Z^1(Y, A)$. Here, $A =G/\bbz_2 =\bbz_2\times \bbz_2$ and $BA$ is the classifying space of $A$.  

Since different extensions are in one-to-one correspondence with the classes in $H^2(\bbz_2 \times \bbz_2, \bbz_2)$, there are $|H^2(\bbz_2 \times \bbz_2, \bbz_2)| =8$ inequivalent extensions \cite{dummit2004abstract}: one for $G =\bbz_2 \times \bbz_2 \times \bbz_2$; one for the quaternion group $G=Q_8 :=\{a, b| a^4 =1, a^2=b^2, ba =a^{-1}b\}$; three for $G =\bbz_4 \times \bbz_2$; and three for $G =D_8 :=\{ a, b| a^2 =b^4=(ab)^2=1\}$. By the K\"{u}nneth theorem \cite{hatcher2002algebraic}, 
$ H^2(\bbz_2 \times \bbz_2, \bbz_2) =\sum_{i =0}^2  H^i(\bbz_2, \bbz_2) \otimes H^{2-i}(\bbz_2, \bbz_2)$. If we denote $g =(g_1, g_2)$, then $e(g)$ is generated by $g_1 \cup  g_1$, $g_1 \cup  g_2$, and $g_2 \cup g_2$ with $\bbz_2$ coefficients. Indeed, there are 8 different elements, each of which corresponds to an extension. The correspondences are listed here: 
\begin{center}
  \begin{tabular}{| l |  r| }
    \hline
    $\bbz_2 \times \bbz_2 \times \bbz_2$  &  0 \\ \hline
    $\bbz_4 \times \bbz_2$ & $g_1\cup g_1$,\ \    $g_1\cup g_1 +g_2\cup g_2$,\ \ $g_2\cup g_2$ \\ \hline
    $D_8$ & $g_1\cup g_1 +g_1\cup g_2$,\ \  $g_1\cup g_2$, \ \  $g_1\cup g_2 +g_2\cup g_2$ \\ \hline
    $Q_8$ & $g_1 \cup  g_1 +g_1 \cup  g_2 +g_2 \cup g_2$  \\
    \hline
  \end{tabular}
\end{center}
The three inequivalent extensions corresponding to  $\bbz_4 \times \bbz_2$ or $D_8$ are obtained by shifting the basis of $\bbz_2 \times \bbz_2$. We can obtain a cocycle representation for $e(g)$ associated with the short exact sequence (\ref{seq1}) from that for $D_8$ by setting $g_2$ in $g_1\cup g_1 +g_1\cup g_2$ to zero. Then we have $g_1\cup g_1$, which also corresponds to $\bbz_4 \times \bbz_2$. In the case of $G= \bbz_4$, we also have $g_1\cup g_1 = \delta \tilde{g}_1/2 \mod 2$, where $\tilde{g}_1$ is    a cochain valued in $\bbz$ that is closed modulo 2. This is because $\beta: H^1(Y, \bbz_2) \to H^2(Y, \bbz_2 \times \bbz_2)$ defined by  $\beta(g_1): = g_1\cup g_1$ can be shown to be equivalent to the connecting (Bockstein) homomorphism induced by the short exact sequence (\ref{seq1}) \cite{hatcher2002algebraic}.

 \bibliographystyle{jhep}
\bibliography{bc} 

\end{document}